\documentclass[a4paper,paper]{JHEP3}

\usepackage{epsfig,multicol}

\title{$\alpha'^2$-corrections to extremal dyonic black holes in heterotic
       string theory}

\author{Maro Cvitan$^{a,c}$, Predrag Dominis Prester$^{b,c}$ and
        Andrej Ficnar$^{c}$\\
   $^a$ International School for Advanced Studies (SISSA/ISAS)\\
        Via Beirut 2--4, 34014 Trieste, Italy\\
   $^b$ Department of Physics,
        University of Rijeka,
        Omladinska 14, HR-51000, Croatia\\
   $^c$ Theoretical Physics Department, Faculty of Science,
        University of Zagreb\\
        p.p. 331, HR-10002 Zagreb, Croatia\\
        E-mail: \email{cvitan@sissa.it}, \email{pprester@ffri.hr},
        \email{aficnar@fizika.org}} 

\abstract{We calculate $\alpha'^2$-corrections to the entropy of the 
5-dimensional 3-charge and the 4-dimensional 4-charge large extremal black 
holes using the low energy effective action of the heterotic string theory. In the 
4-dimensional case, our results are in agreement with the microscopic statistical 
entropy both for the BPS and the non-BPS black holes. In the more interesting 
5-dimensional case, where the direct microscopic stringy description is still 
unknown, our results for the BPS black holes are in agreement with the results
obtained from the action supplemented with $R^2$-correction obtained by 
supersymmetric completion of the gravitational Chern-Simons term. This agreement 
does \emph{not} extend to the non-BPS black holes, for which we propose a different 
expression for the entropy. We show that the new expression is supported by certain 
$\alpha'^3$-order calculations, and by the arguments based on the AdS/CFT 
correspondence.}

\begin{document}

\section{Introduction: Motivation and results}
\label{sec:intro}

Studies of stringy $\alpha'$-corrections to the entropy of black holes
have played an important role in recent years. On  
one hand, conjectures on 
microscopic descriptions of black holes as some multiplets of states in 
string theory were directly tested. On the other hand, these studies
improved our understanding of some concepts, such as the attractor mechanism,
the $AdS_3/CFT_2$ conjecture and dimensional lifts, while also uncovering 
some
interesting relations between black holes and topological strings. Recent 
reviews of these developments can be found in 
\cite{Sen:2007qy,Mohaupt:2007mb,Kraus:2006wn,Larsen:2006xm,%
Guica:2007wd,Pioline:2006ni}.

In this paper we shall deal with two of the simplest cases in their
respective dimensions -- extremal spherically symmetric large black holes 
of the heterotic string theory,  either with three charges in five dimensions, 
or with four charges in four dimensions.

Let us first recapitulate the situation for 4-dimensional 4-charge black holes 
present in the heterotic string theory compactified on $K3\times S^1\times 
\widetilde{S}^1$ or $T^4\times S^1\times \widetilde{S}^1$ background 
with $N$ Kaluza-Klein and $W$ $H$-monopoles wound around the circle 
$\widetilde{S}^1$. If we focus on the states with non-vanishing momentum 
number $n$ and winding number $w$ on the circle $S^1$, for some choices of
relative signs of the charges (e.g., $n,w,N,W$, all positive) these states
are BPS. It is possible to calculate the statistical entropy, i.e. the number 
of such states in the limit of small string coupling constant $g_s$ 
(free string limit), which for $nw\gg1$ is given by\footnote{For the sake 
of clarity we restrict ourselves to the case $w>0$, $NW\ge0$ 
(generalization to other choices of signs is trivial).} 
\cite{Jatkar:2005bh,David:2006ji,David:2006yn}
\begin{equation} \label{A3susyb}
\mathcal{S}_{stat}^{(BPS)} = 2\pi \sqrt{nw(NW+4)} \,, \qquad n>0 \,.
\end{equation}
For $n<0$ the corresponding states are non-BPS with the statistical entropy given by
\begin{equation} \label{A3susyn}
\mathcal{S}_{stat}^{(n-BPS)} = 2\pi \sqrt{|n|w(NW+2)} \,, \qquad n<0
\,.
\end{equation}
Note that (\ref{A3susyb}) and (\ref{A3susyn}) are exact in $\alpha'$.
Now, when one increases $g_s$, it has been argued that at some point 
these states become black holes. While in this regime string theory 
becomes highly nonperturbative, it is expected that one can use low 
energy effective action (at least for large black holes). Indeed, in 
the lowest order in $\alpha'$, the solutions which describe extremal black
holes with the two electric ($n$ and $w$) and the two magnetic charges ($N$ 
and $W$) were explicitly constructed \cite{Cvetic}. The near-horizon effective 
string coupling constant is proportional to $1/|nw|$, which means that
one can neglect string loops for $nw\gg1$. Also, the expansion in 
$\alpha'$ is equivalent to the expansion in $1/|NW|$. The 
Bekenstein-Hawking entropy is $S_{bh(0)}=2\pi\sqrt{|nwNW|}$, in 
agreement with (\ref{A3susyb}) and (\ref{A3susyn}). The $\alpha'$-corrections to the entropies were calculated in 
\cite{Sahoo:2006pm}, with the results again in agreement with (\ref{A3susyb}) and (\ref{A3susyn}). 

Surprising results were obtained when the following two types of 
$R^2$-corrections in the effective action were taken: 
(i) the supersymmetrized gravitational Chern-Simons term 
\cite{LopesCardoso:2004xf}, (ii) the Gauss-Bonnet term \cite{Sen:2005iz}. 
Both of these actions give the black hole entropy in the BPS 
case in the \emph{exact} agreement with the statistical entropy 
(\ref{A3susyb}), while they do not reproduce (\ref{A3susyn}) in the non-BPS 
case. These results are surprising because the full effective action 
contains an infinite number of additional terms, for which there is no 
obvious reason to produce a canceling contribution. Using $AdS_3$-based
arguments, in \cite{Kraus:2005vz,David:2007ak} it was shown that only
effective 3-dimensional gravitational Chern-Simons terms are important
for the calculation of the black hole entropy, and that in this way 
one indeed obtains exactly (\ref{A3susyb}) and (\ref{A3susyn}). This
gives a partial explanation for the success of the action with 
correction (i) (it is not clear why it is not working for non-BPS 
black holes), but the success of the pure Gauss-Bonnet correction remains 
a mystery. Let us mention that (\ref{A3susyb}) was also obtained from 
topological string partition function by using the OSV conjecture 
\cite{topstr}.

One way to acquire a better understanding of what is happening is to 
analyze in the same fashion higher dimensional extremal black holes.
It is known that in five dimensions there are simple 3-charge BPS 
black hole solutions of the lowest order (in $\alpha'$ and $g_s$) 
effective heterotic SUGRA action, which are the obvious candidates.
However, we face several problems here. On the string side,
it is still not known how to calculate the statistical entropy. Also, it is still unknown how to extend the $AdS_3$-based arguments to this
case. On the effective supergravity side, supersymmetrization of the 
5-dimensional gravitational Chern-Simons term was constructed only 
recently in \cite{Hanaki:2006pj}. It was shown in 
\cite{Castro:2007hc,Alishahiha:2007nn,Cvitan:2007en} that the action with
such $R^2$-correction (type (i)) has extremal 3-charge black hole 
solutions with the entropy in the BPS case given by
\begin{equation} \label{Ssusyb}
\mathcal{S}_{bh}^{(BPS)} = 2\pi \sqrt{nw(m+3)}, \qquad n,w>0,\;m\ge0
\end{equation}
while in the non-BPS case we obtain \cite{Cvitan:2007en}
\begin{equation} \label{Ssusyn}
\mathcal{S}_{bh}^{(n-BPS)} = 2\pi \sqrt{|n|w\left(m+\frac{1}{3}\right)}, 
\qquad n<0, \; w,m>0.
\end{equation}
Here $n,w,m$ are integers, with $n$ and $w$ playing the role of 
electric charges and $m$ is the magnetic charge of the 3-form field strength (again, 
for clarity, we restricted ourselves to $w,m>0$).

In the case of the pure Gauss-Bonnet $R^2$-correction (type (ii)) one obtains a more
complicated result for the black hole entropy \cite{Cvitan:2007en}, which has the following expansion in $1/m$ (i.e., in $\alpha'$)
\begin{equation} \label{SGB}
\mathcal{S}_{bh} = 2\pi \sqrt{|nwm|} \left(
 1 + \frac{3}{2|m|} - \frac{3}{4|m|^2} + O\left(m^{-3}\right) \right)
 \,.
\end{equation}

Comparison of (\ref{Ssusyb}) and (\ref{Ssusyn}) with (\ref{SGB}) 
obviously shows that in five dimensions actions with the $R^2$-corrections
of type (i) and (ii) give different results for the black hole 
entropy, which start to differ at the order $\alpha'^2$ for the BPS black 
holes (and already at the order $\alpha'$ for the non-BPS).
It is still unclear, which one, if any, would be expected to agree 
with the (still unknown) statistical entropy of string states. Let us 
mention that it was eventually shown \cite{Castro:2007ci}
(after some initial confusion), that for the BPS black holes it is the
supersymmetric result (\ref{Ssusyb}) which agrees with the prediction
of the OSV conjecture (properly lifted from $D=4$ to $D=5$).

However, a strange thing happens when one considers small black 
holes, which have $m=0$. In this case, on the microscopic (string) 
side, the corresponding states are simple perturbative states, known as the 
Dabholkar-Harvey states, for which the statistical entropy in the BPS 
case is given by \cite{Dabholkar:2004yr,Huang:2007sb}
\begin{equation} \label{Ssmallb}
\mathcal{S}_{stat}^{(BPS)} = 4\pi \sqrt{nw} \,, \qquad n>0 \,,
\end{equation}
and in the non-BPS case by
\begin{equation} \label{Ssmalln}
\mathcal{S}_{stat}^{(n-BPS)} = 2\sqrt2\pi \sqrt{|n|w} \,, \qquad n<0
\,.
\end{equation}
This is obviously different from (\ref{Ssusyb}) when $m=0$. Interestingly, the 
action with the Gauss-Bonnet correction gives in this case
\begin{equation} \label{GBsmall}
	\mathcal{S}_{bh}^{(BPS)} = 4\pi \sqrt{|nw|} \,,
\end{equation}
which agrees with the statistical entropy in the BPS case (\ref{Ssmallb}). So, one 
truncated action appears to work for large black holes, and the other one for
small black holes. Let us mention that this situation was shown to happen
for a class of black holes. In \cite{Huang:2007sb,Cvitan:2007en} it was 
shown that this generalizes to a larger class of small 5-dimensional black holes,
and also that the success of the Gauss-Bonnet action for small black holes can be 
extended to $D>5$ by including higher extended Gauss-Bonnet densities 
\cite{Prester:2005qs}.

In view of all this, we committed ourselves to perturbatively calculate the entropy of 
the large 5-dimensional 3-charge extremal black holes up to the $\alpha'^2$-order 
using low energy effective action of the heterotic string (which is unambiguously 
known only up to the $\alpha'^1$-order). The main virtue is that this is a 
straightforward calculation giving unambiguous results for corrections to 
the black hole entropy, which can be eventually compared with the microscopic 
ones. Our result for the entropy of the BPS black holes is
\begin{equation} \label{Spertb}
\mathcal{S}_{bh}^{(BPS)} =  2\pi \sqrt{nwm}
 \left( 1 + \frac{3}{2m} - \frac{9}{8m^2} + O\left(m^{-3}\right) \right) ,
 \qquad n,w,m>0\,,
\end{equation}
which is in agreement with the supersymmetric result, i.e., with (\ref{Ssusyb})
 expanded in $1/m$.

For the non-BPS black holes we obtain for the entropy
\begin{equation} \label{Spertn}
\mathcal{S}_{bh}^{(n-BPS)} = 2\pi \sqrt{|n|wm}
 \left( 1 + \frac{1}{2m} - \frac{1}{8\,m^2} + O\left(m^{-3}\right)
 \right) , \qquad n<0, \; w,m>0,
\end{equation}
which obviously disagrees with both SUSY (\ref{Ssusyn}) and Gauss-Bonnet 
(\ref{SGB}) results already at $\alpha'^1$-order. 
Instead, our result (\ref{Spertn}) suggests the following formula
\begin{equation} \label{Segzn}
\mathcal{S}_{bh}^{(n-BPS)} = 2\pi \sqrt{|n|w(m+1)}\, .
\end{equation}
Furthermore, if we take the BPS formula (\ref{Ssusyb}) for granted, then we are
able to show that $\alpha'^3$ term in the non-BPS entropy formula
(\ref{Spertn}) must be $1/(16\,m^3)$, which is again in agreement with the
conjectured expression (\ref{Segzn}). Now, using AdS/CFT arguments, from 
(\ref{Spertb}) and (\ref{Spertn}) one infers that central charges satisfy 
$c_L - c_R = 12w$, which is indeed what is expected \cite{Kraus:2007vu}.

The rest of the paper goes as follows. In section \ref{sec:D6het} we
start from the $\alpha'$-corrected low energy effective action of 
heterotic string in $D=6$ and analyze further compactifications on one
or two circles $S^1$. In section \ref{sec:ef-exp} we review Sen's entropy 
function formalism and write perturbative expansions in $\alpha'$.
Section \ref{sec:3-charge} is the central part of the paper in which
we present the results for the entropies of the 5-dimensional 
3-charge extremal black holes up to $\alpha'^2$-order. In section 
\ref{sec:4-charge} we do the same for the 4-dimensional 4-charge black 
holes, which is an extension of the results from \cite{Sahoo:2006pm} to 
order $\alpha'^2$. Our results agree with the microscopic entropy formulas
both for the BPS and non-BPS black holes. In appendix \ref{sec:appen1} we 
exhibit the relations between the charges which appear in section 
\ref{sec:3-charge} with the standard ones. In appendix 
\ref{sec:appen2} we present explicit expressions for the $\alpha'$-corrections
of the near-horizon solutions. In appendix \ref{sec:appen3} we analyze the 
contributions of $\alpha'^2$-terms from the effective action and outline the 
proofs for the properties we use in sections \ref{sec:ef-exp} and 
\ref{sec:3-charge}.

\section{$D=6$ heterotic effective action}
\label{sec:D6het}

We consider the heterotic string compactified on a $T^4$ (or $K3$) manifold.
There is a consistent truncation in which the bosonic part of the 
6-dimensional low energy effective Lagrangian $\mathcal{L}^{(6)}$ is a
function of the string metric $G^{(6)}_{MN}$, Riemann tensor 
$R^{(6)}_{MNPQ}$, dilaton $\Phi^{(6)}$, 3-form $H^{(6)}_{MNP}$ and the 
covariant derivatives of these fields. $H^{(6)}_{MNP}$ is a gauge 
field strength given by
\begin{equation}
H^{(6)}_{MNP} = \partial_M B^{(6)}_{NP} + \partial_N B^{(6)}_{PM}
 + \partial_P B^{(6)}_{MN} - 3\alpha' \Omega^{(6)}_{MNP} \;.
\end{equation}
The last term, $\Omega^{(6)}_{MNP}$, known as the gravitational
Chern-Simons 3-form, is a function of connection and it introduces
terms in the action which are not manifestly diffeomorphism 
invariant.\footnote{We note that in Ref.\ \cite{Sahoo:2006pm} there is
a wrong sign in Eq.\ (3.24) (which propagates to (3.31), (3.33), (3.34)
and (3.36)). This error gets compensated by another one, a wrong sign
in (3.39), which makes the final expression (3.40) again correct.}

It was shown in \cite{Sahoo:2006pm} that, by introducing an additional 3-form
$K^{(6)}=dC^{(6)}$, the theory can be put in a classically
equivalent form with the Lagrangian given by
\begin{eqnarray}
\sqrt{-G^{(6)}} \widetilde{\mathcal{L}}^{(6)} &=&
 \sqrt{-G^{(6)}} \mathcal{L}^{(6)}
 + \frac{1}{(24\pi)^2} \epsilon^{MNPQRS} K^{(6)}_{MNP} H^{(6)}_{QRS}
\nonumber \\
&& + \frac{3\alpha'}{(24\pi)^2} \epsilon^{MNPQRS} 
 K^{(6)}_{MNP} \Omega^{(6)}_{QRS} \;,
\label{gnewact}
\end{eqnarray}
where now $H^{(6)}_{MNP}$ should not be treated as a gauge strength
but as an auxiliary 3-form. Importance of this transformation is that the
problematic Chern-Simons term is now isolated in a way which will
allow us to turn it into a manifestly covariant form in the backgrounds we
are going to consider. 

The 6-dimensional effective Lagrangian has an infinite expansion in 
$\alpha'$
\begin{equation} \label{lalphex}
\mathcal{L}^{(6)} = \sum_{n=0}^\infty \mathcal{L}^{(6)}_n \;,
\end{equation}
where the two lowest terms, in a suitable field redefinition scheme
\cite{Metsaev:1987zx}, and
using the conventions from \cite{Sahoo:2006pm},\footnote{Which means that
$\alpha'=16$, Newton's constant $G_6=2$, and that the 
antisymmetric tensor density $\epsilon^{MNPQRS}$ is defined by 
$\epsilon^{012345}=1$.} are
\begin{eqnarray}
\mathcal{L}^{(6)}_0 &=& \frac{1}{32\pi} e^{-2\Phi^{(6)}} \left[ R^{(6)} +
 4\left(\partial \Phi^{(6)}\right)^2 - \frac{1}{12}H^{(6)}_{MNP}H^{(6)MNP}
 \right]
\label{L60} \\
\mathcal{L}^{(6)}_1 &=& \frac{1}{16\pi} \, e^{-2\Phi^{(6)}} \, 
 \Bigg[ R^{(6)}_{KLMN} R^{(6)KLMN} 
 - \frac{1}{2} R^{(6)}_{KLMN} H_P^{(6)KL} H^{(6)PMN}
\nonumber \\ 
&& - \frac{1}{8} H_K^{(6)MN} H^{(6)}_{LMN} H^{(6)KPQ} H^{(6)L}_{PQ}
 + \frac{1}{24} H^{(6)}_{KLM} H^{(6)K}_{PQ} H_R^{(6)LP} H^{(6)RMQ}
\Bigg] \,.
\label{L61}
\end{eqnarray}

Our goal is to calculate the $\alpha'^2$ correction to the entropy, for
which one would expect that we need $\mathcal{L}^{(6)}_2$. It is known 
that in some schemes (e.g., manifestly supersymmetric) the bosonic part of
$\mathcal{L}^{(6)}_2$ vanishes, but also that field redefinitions 
generally introduce such terms \cite{Chemissany:2007he}. One example 
is presented in \cite{Gross:1986mw} where the $\alpha'^2$-terms have been 
explicitly calculated, but only up to 4-point. A possible way to obtain 
all terms in the scheme we use would be to start with the manifestly 
supersymmetric scheme and extend the analysis of 
\cite{Chemissany:2007he} to the $\alpha'^2$-order. Fortunately, this long 
and tedious calculation is not necessary. As we shall explain at the end 
of Section \ref{sec:ef-exp} (and, in more detail, in appendix 
\ref{sec:appen3}), the contribution of $\mathcal{L}^{(6)}_2$ to
the $\alpha'^2$-corrections of the entropies for the black holes that we analyze in 
this paper vanishes.

Our interest are black holes in $D=5$ and $D=4$ dimensions, so we 
consider further compactification on $6-D$ circles $S^1$. Using the standard
Kaluza-Klein compactification we obtain $D$-dimensional fields
$G_{\mu\nu}$, $C_{\mu\nu}$, $\Phi$, $\widehat{G}_{mn}$,
$\widehat{C}_{mn}$ and $A_\mu^{(i)}$ ($0\le\mu,\nu\le D-1$,
$D\le m,n\le 5$, $1\le i\le 2(6-D)$): 
\begin{eqnarray} \label{d6dD}
&& \widehat{G}_{mn} = G^{(6)}_{mn}\,, \quad
 \widehat{G}^{mn} = (\widehat{G}^{-1})^{mn} \,, \quad
 \widehat{C}_{mn} = C^{(6)}_{mn}\,,
\nonumber \\
&& A^{(m-D+1)}_\mu = \frac{1}{2} \widehat{G}^{nm} G^{(6)}_{n\mu} \,,
 \quad
 A^{(m-2D+7)}_\mu = \frac{1}{2} C^{(6)}_{m\mu}
 - \widehat{C}_{mn} A^{(n-D+1)}_\mu \, ,
\nonumber \\
&& G_{\mu\nu} = G^{(6)}_{\mu\nu} - \widehat{G}^{mn} G^{(6)}_{m\mu}
 G^{(6)}_{n\nu}\, ,
\nonumber \\
&& C_{\mu\nu} = C^{(6)}_{\mu\nu} - 4 \widehat{C}_{mn}
 A^{(m-D+1)}_\mu A^{(n-D+1)}_\nu
 - 2(A^{(m-D+1)}_\mu A^{(m-2D+7)}_\nu
 - A^{(m-D+1)}_\nu A^{(m-2D+7)}_\mu)
\nonumber \\
&& \Phi = \Phi^{(6)} - \frac{1}{2} \ln \mathcal{V}_{6-D} \,,
\end{eqnarray}
There is also (now auxiliary) field $H^{(6)}_{MNP}$ which produces
$D$-dimensional fields $H_{\mu\nu\rho}$, $H_{\mu\nu m}$, 
$H_{\mu mn}$ and $H_{mnp}$. As the 3-form $H$ will respect the same
symmetries as $K$, to simplify the formulae we shall not write it
explicitly but only introduce it when necessary. 

As in \cite{Sahoo:2006pm}, we take for the circle coordinates 
$0\le x^m<2\pi\sqrt{\alpha'}=8\pi$, so that the volume 
$\mathcal{V}_{6-D}$ is
\begin{equation}
\mathcal{V}_{6-D} = (8\pi)^{6-D} \sqrt{\widehat{G}} \;.
\end{equation}
The gauge invariant field strengths associated with
$A_\mu^{(i)}$ and $C_{\mu\nu}$ are
\begin{equation} \label{eag2a}
F^{(i)}_{\mu\nu} =
 \partial_\mu A^{(i)}_\nu - \partial_\nu A^{(i)}_\mu \, ,
 \qquad 1\le i,j\le 2(6-D) \, ,
\end{equation}
\begin{equation} \label{eag2b}
K_{\mu\nu\rho} = \left( \partial_\mu C_{\nu\rho}
 + 2 A_\mu^{(i)} L_{ij} F^{(j)}_{\nu\rho} \right)
 + \hbox{cyclic permutations of $\mu$, $\nu$, $\rho$} \, ,
\end{equation}
where
\begin{equation} \label{edefl}
L = \pmatrix{ 0 & I_{6-D} \cr I_{6-D} & 0} \, ,
\end{equation}
$I_{6-D}$ being a $(6-D)$-dimensional identity matrix.

For the black holes we are going to be interested in, we shall
have\footnote{This means that the second dualization in
\cite{Sahoo:2006pm} (see Eq. (2.16) there), which introduces the 
scalar $b$, is not necessary.}
\begin{equation}
A_\mu^{(i)} L_{ij} F_{\nu\rho}^{(j)} = 0 \;.
\end{equation}

Normally, the next step would be to perform the Kaluza-Klein reduction on
the 6-di\-men\-sional action to obtain a $D$-dimensional low energy
effective action, which can be quite complicated. In 
\cite{Sahoo:2006pm} a simpler procedure is suggested -- one goes to 
$D$  dimensions just to use the symmetries of the action to construct an 
ansatz for the background ($AdS_2\times S^{D-2}$ in our case) and then
performs an uplift to 6 dimensions (by inverting (\ref{d6dD})) where the 
action is simpler and calculations are easier. We shall follow this
logic here.

\section{Entropy function and its expansion}
\label{sec:ef-exp}

We are interested in the near-horizon behavior of the $D$-dimensional
rotationally invariant extremal black holes. We expect that the metric is 
$AdS_2\times S^{D-2}$, which has $SO(2,1)\times SO(D-1)$ as
an isometry group, and that the whole background respects this
symmetry manifestly (note that the Chern-Simons terms are not manifestly
symmetric, so they have to be additionally manipulated). In this case 
one can apply Sen's entropy function formalism \cite{Sen:2005wa,ent-funct}.

The background consists of the metric
$g_{\mu\nu}$, scalars $\phi_s$, two-forms $F^{I}$, and $(D-2)$-form
$H_{m}$.
It follows from the symmetries that 
\begin{eqnarray} \label{efgen}
&& ds^2 = v_1 \left( -r^2 dt^2 + \frac{dr^2}{r^2} \right)
 + v_2\,d\Omega_{D-2}^2 \nonumber \\
&& \phi_s = u_s \;, \qquad s=1,\ldots,n_s  \nonumber \\
&& F^{I}_{rt} = f^I \;, \qquad i=1,\ldots,n_F
 \nonumber \\
&& H_{m} = h_m \mathbf{\epsilon}_{S} \; \qquad m=1,\ldots,n_H
\end{eqnarray}
where $v_{1,2}$, $u_s$, $f^I$ and $h_m$ are constants, and 
$\mathbf{\epsilon}_{S}$ is an induced volume-form on  unit 
$S^{D-2}$. For $F^{I}$ ($H_{m}$),  which are the gauge field 
strengths, $e^I=f^I$ ($q_m=h_m$) are the electric fields 
(magnetic charges).

The near-horizon properties can be obtained from the entropy function
\begin{equation}
\mathcal{E} = 2\pi \left( q_I \, e^I - f \right) \;,
\end{equation}
where $q_I$ are electric charges, and
\begin{equation}
f = \int_{S^{D-2}} \sqrt{-g} \, \mathcal{L} \,.
\end{equation}
If by \{$\varphi_a\}$ we denote the set of the unknowns in (\ref{efgen})
(excluding the electric and the magnetic charges), then the solutions of equations of 
motion, which we denote by $\{\bar{\varphi}_a\}$, are obtained by 
extremization of the entropy function
\begin{equation} \label{geneom}
0 = \frac{\partial\mathcal{E}}{\partial\varphi_a} \,
 \Bigg|_{\varphi=\bar{\varphi}} \;.
\end{equation}
The value of the entropy function at the extremum is equal to Wald's
definition \cite{Wald} of the black hole entropy\footnote{In 
\cite{Tachikawa:2006sz}  Wald formula was extended to actions containing the 
gravitational Chern-Simons terms.}
\begin{equation} \label{bhgen}
\mathcal{S} = \mathcal{E}(\bar{\varphi}) \;.
\end{equation}

In this paper we are interested in the $\alpha'$-corrections, so we need
expansions such as (\ref{lalphex}). Generally, if the Lagrangian has 
expansion in some parameter $\alpha$, the same is true for the respective 
entropy function
\begin{equation} \label{efalphex}
\mathcal{E}(\varphi) = \sum_{n=0}^\infty
 \alpha^n \mathcal{E}_n(\varphi) \;.
\end{equation}
The regular solutions can also be expanded in the same manner
\begin{equation} \label{solalphex}
\bar{\varphi} = \sum_{n=0}^\infty \alpha^n \bar{\varphi}_n \;.
\end{equation}
Putting (\ref{efalphex}) and (\ref{solalphex}) in (\ref{geneom}) we
obtain:
\begin{eqnarray}
&& 0 = \frac{\partial\mathcal{E}_0}{\partial\varphi^a}\,
  \Bigg|_{\varphi=\bar{\varphi}_0} \equiv \bar{\mathcal{E}}_{0,a}
\label{phi0sol} \\
&& \bar{\varphi}_{1}{}^{a} =
  - \bar{\mathcal{E}}_0^{\;,ab} \bar{\mathcal{E}}_{1,b}
\label{phi1sol} \\
&& \bar{\varphi}_{2}{}^{a} = - \bar{\mathcal{E}}_0^{\;,ab} \left(
  \frac{1}{2}\bar{\mathcal{E}}_{0,bcd} \bar{\varphi}_{1}{}^{c}
  \bar{\varphi}_{1}{}^{d} + \bar{\mathcal{E}}_{1,bc} 
\bar{\varphi}_{1}{}^{c}
  + \bar{\mathcal{E}}_{2,b} \right)
\label{phi2sol} \\
&& \vdots \nonumber
\end{eqnarray}
Indices $,ab\ldots$ denote derivatives, and the bar over the function
means that it is evaluated on the $0^{th}$-order solution $\varphi_0$.
For example,
\begin{equation}
\bar{\mathcal{E}}_{1,ab} \equiv
  \frac{\partial^2\mathcal{E}_1}{\partial\varphi^a\partial\varphi^b}
  \,\Bigg|_{\varphi=\bar{\varphi}_0}\;. 
\end{equation}
Also, $\bar{\mathcal{E}}_0^{\;,ab}$ denotes the matrix inverse of
$\bar{\mathcal{E}}_{0,ab}$.

Finally, we expand  the black hole entropy
\begin{equation} \label{bhalphex}
\mathcal{S}_{bh} = \sum_{n=0}^\infty \alpha^n \mathcal{S}_n \;.
\end{equation}
From (\ref{bhgen})-(\ref{phi2sol}) it follows
\begin{eqnarray}
\mathcal{S}_0 &=& \bar{\mathcal{E}}_0 \label{S0sol} \\
\mathcal{S}_1 &=& \bar{\mathcal{E}}_1 \label{S1sol} \\
\mathcal{S}_2 &=& \frac{1}{2} \bar{\mathcal{E}}_{1,a}
  \bar{\varphi}_{1}{}^{a} + \bar{\mathcal{E}}_2
\label{S2sol} \\
\mathcal{S}_3 &=& \frac{1}{6}\bar{\mathcal{E}}_{0,abc}
  \bar{\varphi}_{1}{}^{a} \bar{\varphi}_{1}{}^{b} \bar{\varphi}_{1}{}^{c}
  + \frac{1}{2}\bar{\mathcal{E}}_{1,ab} \bar{\varphi}_{1}{}^{a}
  \bar{\varphi}_{1}{}^{b} + \bar{\mathcal{E}}_{2,a} \bar{\varphi}_{1}{}^{a}
  + \bar{\mathcal{E}}_3
\label{S3sol} \\
&& \vdots \nonumber
\end{eqnarray}
In our calculations we shall take for the expansion parameter 
$\alpha=\alpha'/16=1$.

Our goal is to calculate the entropy up to $\alpha'^2$-order, and from 
(\ref{S2sol}) it may appear that we need the precise form of 
$\mathcal{L}^{(6)}_2$. In appendix \ref{sec:appen3} we show
that from the field content of the effective action, manifest diffeomorphism 
invariance of $\mathcal{L}^{(6)}_2$, and the symmetries of the 0$^{th}$ order
solutions (geometry locally isomorphic to AdS$_3 \times S^3$) follows that
\begin{equation} \label{ef52}
\bar{\mathcal{E}}_2 = 0 \,.
\end{equation}
In the same way, we have also shown that the last two terms in (\ref{S3sol}) 
depend only on the absolute values of the charges (and not on their signs). This will 
allow us to make some conclusions on the $\alpha'^3$-corrections.

\section{3-charge black holes in $D=5$}
\label{sec:3-charge}

Here we consider the 5-dimensional spherically symmetric 3-charge extremal 
black holes which appear in the heterotic string theory compactified on
$T^4\times S^1$ (or $K3\times S^1$). One can obtain an effective
5-dimensional theory by putting $D=5$ in
(\ref{d6dD}) and taking as non-vanishing only the following fields: string metric
$G_{\mu\nu}$, dilaton $\Phi$, modulus $T=(\widehat{G}_{55})^{1/2}$, gauge 
fields $A_\mu^{(i)}$ ($0\le\mu,\nu\le4$, $1\le i\le 2$), and the 3-form
strength $K_{\mu\nu\rho}$. For extremal black holes we expect $AdS_2\times S^3$ 
near-horizon geometry (\ref{efgen}) which in the present case is 
given by:
\begin{eqnarray} \label{d6d5}
&& ds^2 \equiv G_{\mu\nu} dx^\mu dx^\nu
 = v_1\left(-r^2 dt^2 + {dr^2\over r^2}\right) + v_2 d\Omega_3 \,, 
\nonumber \\
&& F^{(1)}_{rt} = \widetilde{e}_1, \qquad
 F^{(2)}_{rt} = \frac{\widetilde{e_2}}{4} \,, \qquad
K_{234} = \frac{\widetilde{p}}{4} \sqrt{g_3} \,,
\nonumber \\
&& S \equiv e^{-2\Phi} = u_S \,,\qquad T = u_T \,.
\end{eqnarray}
Here $g_3$ is a determinant of the metric on the unit 3-sphere $S^3$
(with coordinates $x^i$, $i=2,3,4$).

We now wish to calculate the entropy function up to  second order in
$\alpha'$. First one makes an uplift of (\ref{d6d5}) to six dimensions 
using
(\ref{d6dD}). One gets
\begin{eqnarray} \label{d5d6}
&& ds_6^2 \equiv G^{(6)}_{MN} dx^M dx^N
 = ds^2 + u_T^2 \left( dx^5 + 2\widetilde{e}_1 rdt \right)^2 \,,
\nonumber \\
&& K^{(6)}_{tr5} = \frac{\widetilde{e}_2}{2} \,, \qquad
 K^{(6)}_{234} = K_{234} = \frac{\widetilde{p}}{4} \sqrt{g_3} \,,
\nonumber \\
&& H^{(6)tr5} = \frac{4 h}{v_1 v_2^{3/2} u_S} \,, \qquad
 H^{(6)234} = - \frac{8 h_2}{v_1 v_2^{3/2} u_S \sqrt{g_3}} \,,
\nonumber \\
&& e^{-2\Phi^{(6)}} =  \frac{u_S}{8\pi\, u_T} \,.
\end{eqnarray}
Here $v_1$, $v_2$, $u_S$, $u_T$, $\widetilde{e}_1$, $\widetilde{e}_2$, $h$
and $h_2$ are unknown variables whose solution is to be found by extremizing
the entropy function.
Normalization for $H$ is taken such that the $0^{th}$-order solution gives
\begin{equation} \label{3cHsol0}
h_{20} = \widetilde{e}_{20} \,, \qquad h_0 = \widetilde{p} \,.
\end{equation}
To calculate the $\alpha'$ corrections to the entropy we follow steps 
described in section \ref{sec:ef-exp}. In $0^{th}$-order we have
\begin{equation} \label{ef3c0}
\mathcal{E}_0 = 2\pi \left[ \widetilde{q}_1\widetilde{e}_1
 + \widetilde{q}_2\widetilde{e}_2
 - \int dx^2 dx^3 dx^4 dx^5 \left(
 \sqrt{-G^{(6)}} \mathcal{L}^{(6)}_0 + \frac{1}{(24\pi)^2}
 \epsilon^{MNPQRS} K^{(6)}_{MNP} H^{(6)}_{QRS}
 \right) \right]
\end{equation}
where $\mathcal{L}^{(6)}_0$ is given in (\ref{L60}). Putting
(\ref{d5d6}) in (\ref{ef3c0}) we obtain
\begin{eqnarray} \label{ef50}
\mathcal{E}_0 &=& 2\pi \left[ \widetilde{q}_1\widetilde{e}_1
 + \widetilde{q}_2\widetilde{e}_2 - \frac{\pi}{16} v_1 v_2^{3/2} u_S
 \left( -\frac{2}{v_1} + \frac{6}{v_2}
 + \frac{2 u_T^2 \widetilde{e}_1^2}{v_1^2}
 + \frac{32\, h_2 (2\widetilde{e}_2 - h_2)}{v_1^2\, u_S^2}
 \right. \right.
\nonumber \\
&& \left. \left. \qquad
 - \frac{8 u_T^2 h (2\widetilde{p} - h)}{v_2^3\, u_S^2} \right)
 \right] \,.
\end{eqnarray}
We separate contributions from the $1^{st}$-order in two parts
\begin{equation} \label{ef51}
\mathcal{E}_1 = \mathcal{E}_1' + \mathcal{E}_1'' \,,
\end{equation}
The first contribution is
\begin{equation} \label{ef3c1p}
\mathcal{E}_1' = - 2\pi \int dx^2 dx^3 dx^4 dx^5 
 \sqrt{-G^{(6)}} \mathcal{L}^{(6)}_1 \,,
\end{equation}
where $\mathcal{L}^{(6)}_1$ is given by (\ref{L61}). Putting
(\ref{d5d6}) in (\ref{ef3c1p}) we obtain
\begin{eqnarray} \label{ef51p}
\mathcal{E}_1' &=& - 2\pi^2 v_1 v_2^{3/2} u_S
 \Bigg[ \frac{1}{2v_1^2} + \frac{3}{2v_2^2}
 - \frac{3\widetilde{e}_1^2 u_T^2}{v_1^3}
 + \frac{11 u_T^4 \widetilde{e}_1^4}{2 v_1^4} 
 - \frac{4 u_T^2 h^2}{v_1 v_2^3 u_S^2}
\nonumber \\
&& \qquad \qquad \qquad
 + \frac{4 u_T^4 h^2 \widetilde{e}_1^2}{v_1^2 v_2^3 u_S^2}
 - \frac{40\, u_T^4 h^4}{v_2^6 u_S^4}
 - \frac{48\, h_2^2}{v_1^2 v_2 u_S^2}
 - \frac{640\, h_2^4}{v_1^4 u_S^4}\Bigg] \,.
\end{eqnarray}
The second contribution in (\ref{ef51}) comes from the Chern-Simons term
\begin{equation} \label{ef3c1pp}
\mathcal{E}_1'' = - \frac{1}{6\pi} \int dx^2 dx^3 dx^4 dx^5
 \epsilon^{MNPQRS} K^{(6)}_{MNP} \Omega^{(6)}_{QRS} \,.
\end{equation}
As already mentioned, this part is not manifestly covariant, so we
cannot straightforwardly plug (\ref{d5d6}) in (\ref{ef3c1pp}). 
Fortunately, our 6-dimensional background is of the type for which one
can apply the strategy used in \cite{Sahoo:2006pm}. 

Notice that the expression for the entropy function, like (\ref{ef3c1pp}) has
the form of some effective 2-dimensional action in $(t,r)$ space. The
idea is to find the covariant form of (\ref{ef3c1pp}) in this 2-dimensional
space. We restrict ourselves to the backgrounds which are obtained by 
Kaluza-Klein compactification on $S^3\times S^1$, but beside this
for the moment we have no other restrictions ((\ref{d5d6}) obviously 
belongs to this class). 

Next, notice that the background (\ref{d5d6}) has a form of a product of 
two 3-dimensional backgrounds, the first one is on $(t,r,x^5)$ space 
and the second one on $(x^2,x^3,x^4)$ space (i.e., $S^3$). We now make
further truncation\footnote{It is generally expected that such 
truncation is consistent.} by considering only configurations which 
respect this product structure, for which (\ref{ef3c1pp}) simplifies
to
\begin{equation} \label{ef3c1ppf}
\mathcal{E}_1'' = - \frac{1}{6\pi} \int dx^2 dx^3 dx^4 dx^5
 \epsilon^{ijk} \epsilon^{abc} \left(
 K^{(6)}_{ijk} \Omega^{(6)}_{abc} - \Omega^{(6)}_{ijk} K^{(6)}_{abc}
 \right) \,,
\end{equation}
where $\{a,b,c\}=\{t,r,5\}$ and $\{i,j,k\}=\{2,3,4\}$, and the
convention for the antisymmetric tensor densities is
\begin{equation}
\epsilon^{tr5} = 1 \,, \qquad \epsilon^{234} = 1 \,.
\end{equation}

In three dimensions it is known \cite{Guralnik:2003we,Sahoo:2006vz} that
for the metrics of the form
\begin{equation} \label{3KK2}
ds^2  = \phi(x)
 \left[ g_{mn}(x) dx^m dx^n + \left(dy + 2A_m(x) dx^m\right)^2 \right]
 \,, 
\end{equation}
where $0\le m,n\le1$, we have (modulo total derivative terms)
\begin{equation} \label{CS3cov}
\epsilon^{\alpha\beta\gamma} \Omega_{\alpha\beta\gamma} = 
 \frac{1}{2} \epsilon^{mn} \left[ R^{(2)} F_{mn}
 + 4 g^{m'p'} g^{q'q} F_{mm'} F_{p'q'} F_{qn} \right] \;,
\end{equation}
where $F_{mn}=\partial_m A_n - \partial_n A_m$, $\epsilon^{mn}$ is
antisymmetric with $\epsilon^{01}=1$, and $R^{(2)}$ is a Ricci
scalar obtained from $g_{mn}$. (\ref{CS3cov}) gives us the desired 
manifestly covariant form (in the reduced 2-dimensional space) for the 
Chern-Simons term.

Now we just have to use (\ref{CS3cov}) in (\ref{ef3c1ppf}). For
$(t,r,x^5)$ subspace by comparing (\ref{d5d6}) with (\ref{3KK2}) we 
obtain
\begin{equation}
g_{mn}(x) dx^m dx^n
 = \frac{v_1}{u_T^2} \left( -r^2 dt^2 + \frac{dr^2}{r^2} \right),
\qquad A_0(x) = \widetilde{e}_1 r \,, \qquad \phi(x)=u_T^2 \,.
\end{equation}
Using this in (\ref{CS3cov}) we get
\begin{equation} \label{CStr5}
\epsilon^{abc} \Omega^{(6)}_{abc}
 = 2 \frac{u_T^2}{v_1} \widetilde{e}_1
 - 4 \frac{u_T^4}{v_1^2} \widetilde{e}_1^3 \,.
\end{equation}
For the 3-sphere the Chern-Simons term vanishes
\begin{equation} \label{CS234}
\epsilon^{ijk} \Omega^{(6)}_{ijk} = 0 \,.
\end{equation}
Using (\ref{CStr5}), (\ref{CS234}) and (\ref{d5d6}) in 
(\ref{ef3c1ppf}) we obtain
\begin{equation} \label{ef51pp}
\mathcal{E}_1'' = - 8\pi^2 \widetilde{p} \left(
 \frac{u_T^2}{v_1} \widetilde{e}_1
 - 2 \frac{u_T^4}{v_1^2} \widetilde{e}_1^3 \right) \,.
\end{equation}

% As we have already showed that $\bar{\mathcal{E}}_2$ vanishes, 
% so 
% \begin{equation} \label{ef52}
% \mathcal{E}_2 = - 2\pi \int dx^2 dx^3 dx^4 dx^5
% \sqrt{-G^{(6)}} \mathcal{L}^{(6)}_2 = 0 \,.
% \end{equation}

We now have all the ingredients to calculate the $\alpha'^2$-corrections to the 
entropy. We just take (\ref{ef50}), (\ref{ef51}), (\ref{ef51p}),
(\ref{ef51pp}) and (\ref{ef52}), and put them into
(\ref{phi0sol})-(\ref{phi1sol}) to get the solutions, and into
(\ref{S0sol})-(\ref{S2sol}) to get the entropy. First of all, we need the
$0^{th}$-order solutions. Using (\ref{ef50}) in (\ref{phi0sol}) we
obtain
\begin{eqnarray} \label{3csol0t}
&& v_{20} = 4 v_{10} = \frac{|\widetilde{q}_2|}{\pi} \,, \qquad
 u_{s0} = \frac{1}{|\widetilde{q}_2|}
 \sqrt{8\pi |\widetilde{q}_1 \widetilde{p}|} \,, \qquad
 u_{T0} = \sqrt{\frac{2}{\pi}
 \left| \frac{\widetilde{q}_1}{\widetilde{p}} \right|} \,,
\nonumber \\
&& \widetilde{e}_{10} = \frac{1}{8\widetilde{q}_1}
 \sqrt{\left| 2\widetilde{q}_1 \widetilde{q}_2 \widetilde{p} \right|}
 \,, \qquad
 \widetilde{e}_{20} = h_{20} = \frac{1}{8\widetilde{q}_2}
 \sqrt{\left| 2\widetilde{q}_1 \widetilde{q}_2 \widetilde{p} \right|}
 \,, \qquad h_0 = \widetilde{p} \,.
\end{eqnarray}
Using this in (\ref{S0sol}) we obtain for the black hole entropy in 
the lowest order
\begin{equation} \label{3cS0t}
\mathcal{S}_0 = \frac{\pi}{\sqrt{2}}
 \sqrt{\left| \widetilde{q}_1 \widetilde{q}_2 \widetilde{p} \right|} \,.
\end{equation}
To make comparison with the results from the literature, we need to express
the charges $(\widetilde{q}_1,\widetilde{q}_2,\widetilde{p})$ in terms of the
integer-valued charges $(n,w,m)$ appearing in string/M-theory. The
fastest way to achieve this is to compare (\ref{3csol0t}) with a
solution obtained from the standard effective action for which this
correspondence is known. This is done in appendix~\ref{sec:appen1} and
the result is
\begin{equation} \label{chnorm}
\widetilde{q}_1 = \frac{n}{2} \,,\qquad \widetilde{q}_2 = - 16\pi m
\,,\qquad \widetilde{p} = - \frac{w}{\pi} \,.
\end{equation}
Using this in (\ref{3cS0t}) we obtain
\begin{equation} \label{3cS0}
\mathcal{S}_0 = 2\pi \sqrt{|nwm|} \,,
\end{equation}
which is a well known result. Putting (\ref{chnorm}) into
(\ref{3csol0t}) we obtain
\begin{eqnarray} \label{3csol0}
&& v_{20} = 4 v_{10} = 16 |m| \,, \qquad
 u_{s0} = \frac{\sqrt{|nw|}}{8\pi |m|} \,, \qquad
 u_{T0} = \sqrt{ \left| \frac{n}{w} \right|} \,,
\nonumber \\
&& \widetilde{e}_{10} = \frac{1}{n} \sqrt{|nwm|} \,, \qquad
 \widetilde{e}_{20} = h_{20} = -\frac{\sqrt{|nwm|}}{32\pi m}
 \,, \qquad h_0 = -\frac{w}{\pi} \,.
\end{eqnarray}

From (\ref{3csol0}) we get the following conclusions. First, to have a 
small near-horizon effective string coupling $g_s^2=1/u_s$, one 
requires $n,w\gg m$. In this regime one can ignore the string loop 
corrections and use the tree level effective action. Second, the Ricci scalar
$R$ and the field strengths $F^2$ and $H^2$ are proportional to $1/m$,
which means that the $\alpha'$ expansion is effectively an expansion in
$1/m$. 

The rest of the procedure is straightforward. As the corrections depend on the
relative signs of charges, we present solutions for two representative
cases:
\begin{itemize}
\item $n,w,m>0$ (BPS solutions),
\item $n<0$, $w,m>0$ (non-BPS solutions).
\end{itemize}
The near-horizon solutions up to $\alpha'^1$-order are presented in
appendix \ref{sec:appen2}. For the entropies we obtain (up to $\alpha'^2$-order):
\begin{eqnarray}
\mathcal{S}_{bh}^{(BPS)} &=& 2\pi \sqrt{nwm} \left(
 1 + \frac{3}{2m} - \frac{9}{8m^2} + O\left(m^{-3}\right)
 \right), \qquad n,w,m>0
\label{Sbps} \\
\mathcal{S}_{bh}^{(n-BPS)} &=& 2\pi \sqrt{|n|wm} \left(
 1 + \frac{1}{2m} - \frac{1}{8m^2} + O\left(m^{-3}\right)
 \right), \quad n<0,\;\; w,m>0
\label{Snbps}
\end{eqnarray}
Comparison with (\ref{Ssusyb}) makes it obvious that for the BPS black holes 
our result (\ref{Sbps}) is in agreement with the result obtained from the 
supersymmetric $R^2$-corrected action (and in disagreement with the
Gauss-Bonnet result (\ref{SGB})). For the non-BPS black holes our result
(\ref{Snbps}) disagrees already at $\alpha'$-order with the results based 
on either SUSY (\ref{Ssusyn}) or Gauss-Bonnet (\ref{SGB}) $R^2$-corrections. 

Observe that (\ref{Snbps}) suggests the following formula
\begin{equation} \label{Snconj}
\mathcal{S}_{bh}^{(n-BPS)} = 2\pi \sqrt{|n|w(m+1)} 
\,\qquad n<0,\;w,m>0 \,.
\end{equation}
This is further supported by the following higher-order arguments.

Using (\ref{S3sol}) we can calculate the $\alpha'^3$-corrections of the 
entropy, with the result
\begin{eqnarray}
\mathcal{S}_3^{(BPS)} &=& 2\pi\sqrt{nwm}\,\frac{571}{16}\frac{1}{m^3}
 + \bar{\mathcal{E}}_{2,a} \bar{\varphi}_{1}{}^{a} + \bar{\mathcal{E}}_3
 \,, \qquad\quad n,m,w>0 
 \label{Sb3} \\
\mathcal{S}_3^{(n-BPS)} &=& 2\pi\sqrt{|n|wm}\,\frac{545}{16}\frac{1}{m^3}
 + \bar{\mathcal{E}}_{2,a} \bar{\varphi}_{1}{}^{a} + \bar{\mathcal{E}}_3
 \,, \qquad n<0, \;\; w,m>0 \,.
\label{Sn3}
\end{eqnarray}
To calculate $\bar{\mathcal{E}}_{2,a}$ and $\bar{\mathcal{E}}_3$ one needs the 
precise knowledge of $\alpha'^2$ and $\alpha'^3$ ($R^4$) parts of the effective 
heterotic action, which is unknown. But, as we explain in appendix 
\ref{sec:appen3}, it can be shown that 
$\bar{\mathcal{E}}_{2,a} \bar{\varphi}_{1}{}^{a}$ and $\bar{\mathcal{E}}_3$ do
not depend on sign assignments for the charges. This means that the last two terms
in (\ref{Sb3}) and (\ref{Sn3}) are equal. Now, if the BPS entropy formula 
(\ref{Ssusyb}) is correct (at least up to 3$^{rd}$-order), from (\ref{Sb3}) we 
obtain 
\begin{equation}
\bar{\mathcal{E}}_{2,a} \bar{\varphi}_{1}{}^{a} + \bar{\mathcal{E}}_3
 = - 2\pi\sqrt{|n|wm}\,\frac{544}{16}\frac{1}{m^3} \,.
\end{equation}
Using this in
(\ref{Sn3}) gives us
\begin{equation}
\mathcal{S}_3^{(n-BPS)} = 2\pi\sqrt{|n|wm}\frac{1}{16}\frac{1}{m^3} \,,
\end{equation}
which is again in agreement with (\ref{Snconj}).

One can extend this argument to all orders using AdS$_3$
argumentation. The AdS/CFT conjecture says that the black hole entropy is equal
to the microcanonical entropy of the boundary 2D CFT, which is given by the
Cardy formula \cite{cardy}. In our case one obtains
\begin{eqnarray*}
\mathcal{S}_{CFT}^{(BPS)} &=& 2\pi\sqrt{\frac{c_L n}{6}} \,\qquad n>0
\\
\mathcal{S}_{CFT}^{(n-BPS)} &=& 2\pi\sqrt{\frac{c_R |n|}{6}} \,\qquad
n<0
\end{eqnarray*}
where $c_L$ ($c_R$) is the central charge of the left (right)
Virasoro algebra. In \cite{Kraus:2007vu} it was shown that in our case
one expects $c_L-c_R=12w$. This is exactly what follows from
(\ref{Ssusyb}) and (\ref{Snconj}). In summary, this argument shows
that if (\ref{Ssusyb}) is correct (and there are reasons, explained in the
introduction, to believe that it is), then (\ref{Snconj}) is also correct. Our
explicit perturbative calculation then reinforces a belief that both
(\ref{Ssusyb}) and (\ref{Snconj}) are correct.

\section{4-charge black holes in $D=4$}
\label{sec:4-charge}

Here we consider the 4-dimensional 4-charge extremal black holes 
appearing in the heterotic string theory compactified on 
$T^4\times S^1\times\tilde{S}^1$ (or $K3\times S^1\times\tilde{S}^1$). 
One can obtain an effective 4-dimensional theory by putting $D=4$ in
(\ref{d6dD}) and taking as non-vanishing only the following fields: string 
metric $G_{\mu\nu}$, dilaton $\Phi$, moduli 
$T_1=(\widehat{G}_{44})^{1/2}$ and $T_2=(\widehat{G}_{55})^{1/2}$, 
and the gauge fields $A_\mu^{(i)}$ ($0\le\mu,\nu\le3$, $1\le i\le 4$). The black hole 
is charged purely electrically with respect
to $A_\mu^{(1)}$ and $A_\mu^{(3)}$, and purely magnetically with
respect to $A_\mu^{(2)}$ and $A_\mu^{(4)}$. Again, for extremal black holes one
expects $AdS_2\times S^2$ near-horizon geometry (\ref{efgen}) which in the
present case is given by:
\begin{eqnarray} \label{d6d4}
&& ds^2 \equiv G_{\mu\nu} dx^\mu dx^\nu
= v_1\left(-r^2 dt^2 + {dr^2\over r^2}\right) +
v_2(d\theta^2 + \sin^2\theta d\phi^2)\, , \nonumber \\
&& e^{-2\Phi}=u_S \,,\qquad T_1=u_1 \,,\qquad T_2=u_2
\nonumber \\
&& F^{(1)}_{rt} = \widetilde{e}_1, \qquad F^{(3)}_{rt}
 = \frac{\widetilde{e_3}}{16} \,, \qquad
 F^{(2)}_{\theta\phi} = \frac{\widetilde{p}_2}{4\pi} \sin\theta \,,
 \qquad F^{(4)}_{\theta\phi} = \frac{\widetilde{p}_4}{64\pi}\sin\theta \,.
\end{eqnarray}

One proceeds in the similar fashion as in section \ref{sec:3-charge}. As 
basically all the building blocks were given in
\cite{Sahoo:2006pm,Exirifard:2006wa} (where only $\alpha'$-correction 
to the entropy was calculated), we shall just 
state the results. In this case the $\alpha'$ expansion is an expansion in 
$1/NW$. For clarity we again take two representative cases:
\begin{itemize}
\item $n,w,N,W > 0$ (BPS),
\item $n<0$, $w,N,W > 0$ (non-BPS).
\end{itemize}
The near-horizon solutions are presented in appendix \ref{sec:appen2}.

We obtain for the entropy up to $\alpha'^2$-order
\begin{eqnarray}
\mathcal{S}_{bh}^{(BPS)} &=& 2\pi \sqrt{nwNW} \left(
 1 + \frac{2}{NW} - \frac{2}{(NW)^2} + O\left((NW)^{-3}\right)
 \right) \,, \qquad n>0 \,,
\label{S4bps} \\
\mathcal{S}_{bh}^{(n-BPS)} &=& 2\pi \sqrt{|n|wNW} \left(
 1 + \frac{1}{NW} - \frac{1}{2(NW)^2} + O\left((NW)^{-3}\right)
 \right) \,, \quad n<0 \,.
\label{S4nbps}
\end{eqnarray}
We see that the results agree with the microscopic entropies 
(\ref{A3susyb}) and (\ref{A3susyn}).

We mention that the arguments considering $\alpha'^3$ and higher order 
corrections (presented at the end of section \ref{sec:3-charge}) can be 
repeated here.

\acknowledgments

We would like to thank L.\ Bonora, M.\ Haack, P.\ Kraus and S.\ Pallua for 
stimulating discussions. This work was supported by the Croatian 
Ministry of Science, Education and Sport under the contract No.\ 
119-0982930-1016. P.D.P. was also supported by Alexander von Humboldt
Foundation, and M.C. by Central European Initiative (CEI).

\appendix

\section{Identification of charges}
\label{sec:appen1}

We start from the 5-dimensional effective Lagrangian of the heterotic string 
compactified on $T^5\times S^1$
\begin{equation} \label{a1-lhet0}
\mathcal{L}_0 = \frac{1}{32\pi} e^{-2\Phi}  \left[ R + 4(\partial \Phi)^2
 - \frac{(\partial T)^2}{T^2} - \frac{1}{12}\left(H_{\mu\nu\rho}\right)^2
 - T^2 \left(F^{(1)}_{\mu\nu}\right)^{\!2}
 - \frac{1}{T^2} \left(F^{(2)}_{\mu\nu}\right)^{\!2} \right] \,.
\end{equation} 
We take the $AdS_2\times S^3$ ansatz for the background
\begin{eqnarray} \label{a1-d5}
&& ds^2 = v_1\left(-r^2 dt^2 + {dr^2\over r^2}\right) + v_2 d\Omega_3 \,, 
\nonumber \\
&& F^{(1)}_{rt} = e_1, \qquad
 F^{(2)}_{rt} = e_2 \,, \qquad
H_{234} = p \sqrt{g_3} \,,
\nonumber \\
&& e^{-2\Phi} = u_S \,,\qquad T = u_T \,.
\end{eqnarray}
The entropy function is given by
\begin{equation} \label{a1-ef0}
\mathcal{E}_0 = 2\pi \left[ q_1 e_1 + q_2 e_2 - \frac{\pi}{16} v_1 v_2^{3/2} u_S
\left( \frac{6}{v_2} - \frac{2}{v_1} + \frac{2 u_T^2 e_1^2}{v_1^2} +
\frac{2 e_2^2}{u_T^2 v_1^2} - \frac{p^2}{2 v_2^3} \right) \right] \,.
\end{equation}
The solutions are
\begin{eqnarray} \label{a1-sol0}
&& v_{20} = 4 v_{10} = \frac{|p|}{2} \,, \qquad
 u_{S0} = \frac{8}{\pi |p|} \sqrt{|q_1 q_2|} \,, \qquad
 u_{T0} = \sqrt{\left| \frac{q_1}{q_2} \right|} \,,
\nonumber \\
&& e_{10} = \frac{1}{4\sqrt{2}\, q_1} \sqrt{\left| q_1 q_2\, p \right|} \,, \qquad
 e_{20}  = \frac{1}{4\sqrt{2}\, q_2} \sqrt{\left| q_1 q_2\, p \right|} \,,
\end{eqnarray}
while the entropy is
\begin{equation}
\mathcal{S}_0 = \frac{\pi}{2} \sqrt{2|q_1q_2p|} \,.
\end{equation}

It is known (see e.g., \cite{Sen:2005iz,Prester:2005qs,Cvitan:2007en})
that the relation with the integer-valued charges $(n,w,m)$ of the string 
theory is given by
\begin{equation} \label{a1-chn}
q_1 = \frac{2n}{\sqrt{\alpha'}} = \frac{n}{2} \,, \qquad 
q_2 = \frac{2w}{\sqrt{\alpha'}} = \frac{w}{2} \,, \qquad 
p = 2\alpha'm = 32\,m \,,
\end{equation}
where we used the convention $\alpha'=16$.
Using this in (\ref{a1-sol0}) we obtain for the solutions
\begin{eqnarray} \label{a1-sol0s}
&& v_{20} = 4 v_{10} = 16|m| \,, \qquad
 u_{S0} = \frac{\sqrt{|nw|}}{8\pi |m|} \,, \qquad
 u_{T0} = \sqrt{\left| \frac{n}{w} \right|} \,,
\nonumber \\
&& e_{10} = \frac{1}{n} \sqrt{|nwm|} \,, \qquad
 e_{20}  = \frac{1}{w} \sqrt{|nwm|} \,,
\end{eqnarray}
and for the entropy a well-known result
\begin{equation}
\mathcal{S}_0 = 2\pi \sqrt{|nwm|} \,.
\end{equation}
Now, by comparing the expressions for $v_{10}$, $u_{S0}$ and $u_{T0}$ in 
(\ref{a1-sol0s}) and (\ref{3csol0t}) one immediately obtains (\ref{chnorm}) 
up to signs. To get the correct signs one has to compare the expressions for the field 
strengths.

First notice that the gauge field $A^{(1)}$ was not involved in 
transformations made in section \ref{sec:D6het}, so 
$\widetilde{e}_1=e_1$ and
\begin{equation}
\widetilde{q}_1 = q_1 = \frac{n}{2} \;,
\end{equation}
where we used (\ref{a1-chn}).
From (\ref{a1-d5}) and (\ref{a1-sol0}) we get
\begin{equation}
H^{234} = \frac{8\,p}{|p|^3 \sqrt{g_3}} \;,
\end{equation}
while from (\ref{d5d6}) and (\ref{3csol0t}) we get
\begin{equation}
H^{234} = H^{(6)234}
 = -\frac{2\pi^2\widetilde{q}_2}{|\widetilde{q}_2|^3 \sqrt{g_3}} \;.
\end{equation}
By comparing the two results and using (\ref{a1-chn}) we obtain
\begin{equation}
\widetilde{q}_2 = -\frac{\pi}{2}p = -16\pi m \;.
\end{equation}
In a similar fashion, by studying $H^{015}$ we finally obtain
\begin{equation}
\widetilde{p} = -\frac{2}{\pi} q_2 = -\frac{w}{\pi} \;,
\end{equation}
which completes the identification (\ref{chnorm}).

\section{Near-horizon solutions}
\label{sec:appen2}

Here we present explicitly $\alpha'$-corrections of the near-horizon solutions 
of the extremal black holes analyzed in the paper. They are obtained from 
(\ref{solalphex})-(\ref{phi1sol}).

\subsection{$D=5$ 3-charge extremal black holes} 

For $n,w,m>0$ (BPS case):
\begin{eqnarray} \label{sol5bv1}
v_1&=&4m\left(1 + O(m^{-2})\right)\\
v_2&=&16m\left(1 + \frac{2}{m} + O(m^{-2})\right)\\
u_S&=&\frac{\sqrt{nw}}{8\pi m}\left(1 - \frac{5}{2m} + O(m^{-2})\right)\\
u_T&=&\sqrt{\frac{n}{w}}\left(1 - \frac{3}{2m} + O(m^{-2})\right)\\
\widetilde{e}_1&=&\frac{1}{n}\sqrt{nwm}\left(1 + \frac{3}{2m}
 + O(m^{-2})\right)\\
\widetilde{e}_2&=&-\frac{\sqrt{nwm}}{32\pi m}\left(1 - \frac{3}{2m}
 + O(m^{-2})\right)\\
h&=&-\frac{w}{\pi}\left(1 + \frac{4}{m} + O(m^{-2})\right)\\
h_2&=&-\frac{\sqrt{nwm}}{32\pi m}\left(1 - \frac{11}{2m} + O(m^{-2})\right)
\label{sol5bh2}
\end{eqnarray}
while for $n<0$, $w,m>0$ (non-BPS case):
\begin{eqnarray} \label{sol5nv1}
v_1&=&4m\left(1 - \frac{12}{m^2} + O(m^{-2})\right)\\
v_2&=&16m\left(1 + \frac{2}{m} + O(m^{-2})\right)\\
u_S&=&\frac{\sqrt{|n|w}}{8\pi m} \left(1 - \frac{3}{2m} + O(m^{-2})\right)\\
u_T&=&\sqrt{\frac{|n|}{w}}\left(1 - \frac{1}{2m} + O(m^{-2})\right)\\
\widetilde{e}_1&=&\frac{1}{n}\sqrt{|n|wm} \left(1 + \frac{1}{2m}
 + O(m^{-2})\right)\\
\widetilde{e}_2&=&-\frac{\sqrt{|n|wm}}{32\pi m} \left(1 -\frac{1}{2m}
 + O(m^{-2})\right)\\
h&=&-\frac{w}{\pi} \left(1 + \frac{4}{m} + O(m^{-2})\right)\\
h_2&=&-\frac{\sqrt{|n|wm}}{32\pi m} \left(1 - \frac{9}{2m}
 + O(m^{-2})\right)
\label{sol5nh2}
\end{eqnarray}

\subsection{$D=4$ 4-charge extremal black holes} 

For $n,w,N,W>0$ (BPS case):
\begin{eqnarray}
v_1 &=& 4NW \left( 1 + \frac{1}{NW}  + O((NW)^{-2})\right)\\
v_2 &=& 4NW \left( 1 + \frac{3}{NW} + O((NW)^{-2})\right)\\
u_S &=& \sqrt{\frac{nw}{NW}} \left( 1 - \frac{2}{NW} + O((NW)^{-2})\right)\\
u_1 &=& \sqrt{\frac{n}{w}} \left( 1 - \frac{3}{2NW} + O((NW)^{-2} ) \right)\\
u_2 &=& \sqrt{\frac{W}{N}} \left( 1 + \frac{3}{2NW} + O((NW)^{-2} ) \right)\\
\widetilde{e}_1 &=& \frac{1}{n}\sqrt{nwNW} \left( 1 + \frac{2}{NW}
 + O((NW)^{-2})\right)\\
\widetilde{e}_3 &=& - \frac{\sqrt{nwNW}}{8\pi W} \left( 1 - \frac{2}{NW}
 + O((NW)^{-2})\right)\\
h_3 &=& - \frac{\sqrt{nwNW}}{8\pi W} \left( 1 - \frac{6}{NW}
 + O((NW)^{-2})\right)\\
h_4 &=& - \frac{w}{2} \left( 1 + \frac{4}{NW} + O((NW)^{-2})\right)
\end{eqnarray}
while for $n<0$, $w,N,W>0$ (non-BPS case):
\begin{eqnarray}
v_1 &=& 4NW \left( 1 + \frac{1}{NW} + O((NW)^{-2})\right)\\
v_2 &=& 4NW \left( 1 + \frac{3}{NW} + O((NW)^{-2})\right)\\
u_S &=& \sqrt{\frac{|n|w}{NW}} \left( 1 - \frac{1}{NW} + O((NW)^{-2})\right)\\
u_1 &=& \sqrt{\frac{|n|}{w}} \left( 1 - \frac{1}{2NW} + O((NW)^{-2})\right)\\
u_2 &=& \sqrt{\frac{W}{N}} \left( 1 + \frac{3}{2NW} + O((NW)^{-2} ) \right)\\
\widetilde{e}_1 &=& \frac{1}{n}\sqrt{|n|wNW} \left( 1 + \frac{1}{NW}
 + O((NW)^{-2})\right)\\
\widetilde{e}_3 &=& - \frac{\sqrt{|n|wNW}}{8\pi W} \left( 1 - \frac{1}{NW}
 + O((NW)^{-2})\right)\\
h_3 &=& - \frac{\sqrt{|n|wNW}}{8\pi W} \left( 1 - \frac{5}{NW}
 + O((NW)^{-2})\right)\\
h_4 &=& - \frac{w}{2} \left( 1 + \frac{4}{NW} + O((NW)^{-2})\right)
\end{eqnarray}
Variables $h_3$ and $h_4$ are here introduced in $H^{(6)}$ in an analogous way 
as $h_2$ and $h$ were in (\ref{d5d6}) and (\ref{3cHsol0}), meaning that at 
$0^{th}$ order they give
\begin{equation} 
h_{30} = \widetilde{e}_{30} \,, \qquad h_{40} = \widetilde{p}_4
\end{equation}

\section{On contributions from $\alpha'^2$ and higher order terms in the action}
\label{sec:appen3}

In our calculations we needed contributions coming from the $\alpha'^2$ 
(six-derivative) and $\alpha'^3$ (eight-derivative) sectors of the 6-dimensional 
heterotic effective action, which still have not been obtained in a direct 
manner.
%\footnote{Contributions from $\alpha'^2$-terms could trivially be 
%obtained by using manifestly supersymmetric action obtained by 
%supersymmetrization of gravitational Chern-Simons term, in which bosonic 
%six-derivative terms are not to be present. As we want to ... }
More precisely, we need: (i) $\bar{\mathcal{E}}_2$, (ii) a difference between the
BPS and the non-BPS results for $\bar{\mathcal{E}}_{2,a} \bar{\varphi}_{1}{}^{a}$,
and (iii) the same for $\bar{\mathcal{E}}_3$ (for notation see section 
\ref{sec:ef-exp}). We shall now show that all these quantities vanish, by
using the following properties: 
\begin{enumerate}
	\item[(1)]  Manifest diffeomorphism covariance (once we 
have isolated the Chern-Simons term to appear only in $\mathcal{L}_1'$, all 
other $\mathcal{L}_n$ are scalars built from the metric, Riemann tensor, 3-form 
field $H$, and the covariant derivatives of them and of dilaton).
\item[(2)] Properties of the near-horizon background ($\nabla_\mu S$ and 
 $\nabla_\mu H_{\nu\rho\sigma}$ vanish).
\item[(3)] The 0$^{th}$-order solution is locally isomorphic to AdS$_3 \times S^3$ 
 (implying that all the covariant derivatives of the Riemann tensor vanish).
\item[(4)] Evaluated on the 0$^{th}$-order solutions we have 
	$R_{\mu\nu\rho\sigma} = H_{\mu\nu}{}^{\tau} H_{\tau\rho\sigma}/4$, and 
(in the vielbein basis)
 $|H_{015}| = |H_{234}|$, 
\item[(5)] If one defines a $L$-derivative as an action of an  operator $L$ (we
 specialize here to 5-dimensional black holes, extension to the 4-dimensional
 case is straightforward)
\begin{equation}
L \equiv e_1 \frac{d}{de_1} - e_2 \frac{d}{de_2} - h_2 \frac{d}{dh_2}
 - S \frac{d}{dS} - T \frac{d}{dT},
\end{equation}
 followed by the substitution of the 0$^{th}$-order solution (\ref{3csol0}), it can
 be shown that the $L$-derivative of vielbein basis components of Riemann, 3-form $H$, and of 
 covariant derivatives of Riemann vanishes.
\end{enumerate}

Let us first consider quantities $\bar{\mathcal{E}}_n$ (excluding 
$\bar{\mathcal{E}}_1'$ which contains the Chern-Simons term). From (3) and (4) it 
follows that every monomial which appears in $\bar{\mathcal{E}}_n$ is equal 
to a constant times a monomial consisting only of $H$-fields (more 
precisely, $2(n+1)$ of them). From (3), it is
easy to see that such monomial is an even function of the field strengths. A
consequence is that $\bar{\mathcal{E}}_n$ do not depend on the signs of 
charges, and consequently give the same result for our BPS and non-BPS
solutions. The special case $n=3$ then settles (iii). 

From (4) it also follows that every monomial is, up to a numerical 
constant, given by $|H_{234}|^{2(n+1)} (1-(-1)^n)$, from which it follows that 
\begin{equation} \label{bEn}
\bar{\mathcal{E}}_n = 0 \qquad \mbox{for $n$ even.}
\end{equation}
For the special case $n=2$ this gives (\ref{ef52}).

Now, we establish that for even $n$, 
$\bar{\mathcal{E}}_{n,a} \bar{\varphi}_{1}{}^{a}$ gives the same result for
 our BPS and non-BPS solutions (given in 
(\ref{sol5bv1}-\ref{sol5bh2}) and (\ref{sol5nv1}-\ref{sol5nh2}), respectively),
i.e.\ it does not depend on the signs of the charges.
Notice that this will be the case if the  $L$-derivative, defined above in (5), 
vanishes when acting on $\mathcal{E}_n$. 
Because of the property (5) we have  
$L(\mathcal{E}_n) \propto L(\sqrt{-G}\,S) \bar{\mathcal{L}}_n$. 
Analogously to (\ref{bEn}) we finally get
\begin{equation} \label{LEn}
L(\mathcal{E}_{n}) = 0 
\qquad \mbox{for $n$ even.}
\end{equation}
Taking the special case $n=2$ settles (ii) and concludes our proof.

\end{document}